RESEARCH ARTICLE                                                                                          OPEN ACCESS

# A Survey of Digital Watermarking Techniques and its Applications

Lalit Kumar Saini[1], Vishal Shrivastava[2]
M.Tech[1] Research Scholar, Professor[2]
Department of Computer Science and Engineering,
Arya College of Engineering. & Information Technology,
Jaipur, India

**ABSTRACT**
Digital media is the need of a people now a day as the alternate of paper media. As the technology grown up digital media required protection while transferring through internet or others mediums. Watermarking techniques have been developed to fulfill this requirement. This paper aims to provide a detailed survey of all watermarking techniques specially focuses on image watermarking types and its applications in today's world.
*Keywords: -* Watermarking, DWT, Copyright, PSNR. .

## I. INTRODUCTION

Today's generation is witness of developments of digital media. A very simplest example of digital media is a photo captured by phone camera. The use of Digital media is common in present era. Other example of Digital media is text, audio, video etc.

We know an internet is the fastest medium of transferring data to any place in a world. As this technology grown up the threat of piracy and copyright very obvious thought is in owners mind. So Watermarking is a process of secure data from these threats, in which owner identification (watermark) is merged with the digital media at the sender end and at the receiver end this owner identification is used to recognize the authentication of data. This technique can be applied to all digital media types such as image, audio, video and documents. From many years researchers and developers worked in this area to gain best results.

 The paper is organized as follows sections:

* Overview of Image watermarking including history of watermarking
* Types of Image watermarking techniques in detail
* Classification & Applications of  watermarking
* Threats for Image watermarking

## II. GENERAL INTRODUCTION OF IMAGE WATERMARKING

Image Watermarking is the technique of embedding of owner copyright identification with the host image. When and how watermarking is used first is the topic of discussion but it can used at Bologna, Italy in 1282 .at first it is used in paper mills as paper mark of company [1]. Then it is common in practice up to 20<sup>th</sup> century. After that watermark also used in the postage stamp and currency notes of any country.
Digital image watermarking is actually derive from Steganography, a process in which digital content is hide with the other content for secure transmission of Digital data. In particular conditions steganography and watermarking are very similar when the data to be secure is hidden in process of transmission over some carrier.
The main difference between these two processes is in steganography the hidden data is on highest priority for sender and receiver but in watermarking bot source image and hidden image, signature or data is on highest priority.

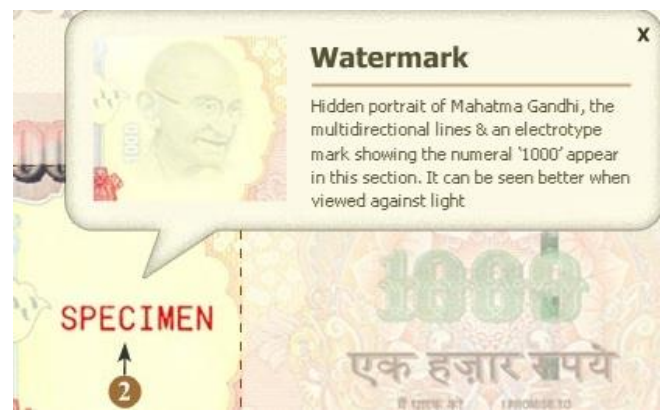

Fig 1 Example of watermark on Indian currency

*2.1. Process of Image Watermarking*

   The process of watermarking is divided into two parts:
   a) Embedding of watermark into host image.
   b) Extraction of watermark from image.

*2.1.1. Watermarking Embedding*

The process of image watermarking is done at the source end. In this process watermark is embedding in the host image by using any watermarking algorithm or process. The whole process is shown in figure 2

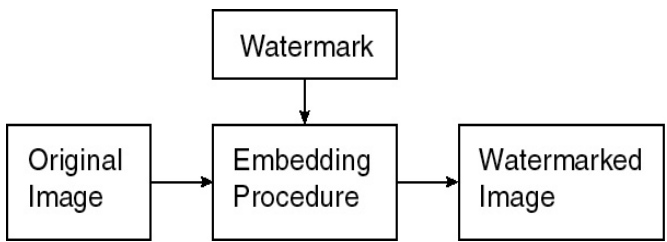

Fig 2 Embedding process of image watermarking

### 2.1.2. Watermarking Extraction

This is the process of Extracting watermark from the watermarked image by reverse the embedding algorithm. The whole process is shown in figure 3

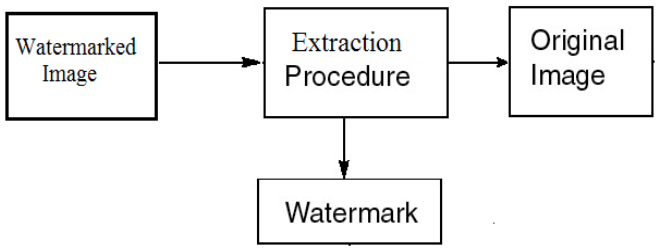

Fig 3 Extraction process of watermark

### 2.2 Watermarking Properties

Watermarking need some desirable properties based on the application of the watermarking system [2].
Some of the properties are presented here:

### 2.2.1. Effectiveness:

This is the most important property of watermark that the watermark should be effective means it should surely be detective. If this will not happened the goal of the watermarking is not fulfilled.

### 2.2.2 Host signal Quality:

This is also important property of watermarking. Everybody knows that in watermarking, watermark is embedded in host signal (image, video, audio etc.). This may put an effect on the host signal. So the watermarking system should be like as, it will minimum changes the host signal and it should be unnoticeable when watermark is invisible.

### 2.3.3. Watermark Size

Watermark is often use to owner identification or security confirmation of host signal and it always use when data is transmitted. So it is important that the size of watermark should be minimum because it will increases the size of data to be transmitted.

### 2.4.4. Robustness

Robustness is crucial property for all watermarking systems. There are so many causes by which watermark is degraded, altered during transmission, attacked by hackers in paid media applications. So watermark should robust, So that it withstand against all the attacks and threats.

## III. CLASSIFICATION & APPLICATIONS

### 3.1 Classification

Digital watermarking techniques are classified into various types. This classification based on several criteria. All the classification describe in following table.

| S.no | Criteria | Classification |
|---|---|---|
| 1. | Watermark Type | 1. Noise: pseudo noise, Gaussian random and chaotic sequences<br>2. Image: Any logo, Stamp Image etc. |
| 2. | Robustness | 1. Fragile: Easily Manipulated.<br>2. Semi-Fragile: Resist from some type of Attacks<br>3. Robust: not affected from attack |
| 3. | Domain | 1. Spatial: LSB, Spread Spectrum<br>2. Frequency: DWT, DCT, DFT, SVD |
| 4. | Perceptivity | 1. Visible Watermarking: Channel logo<br>2. Invisible Watermarking: like Steganography |
| 5. | Host Data | 1. Image Watermarking<br>2. Text Watermarking<br>3. Audio Watermarking<br>4. Video Watermarking |
| 6. | Data Extraction | 1. Blind<br>2. Semi-Blind<br>3. Non- Blind |

**Table 1:** Types of watermarking basis of different Criteria

In the image watermarking domain based techniques is generally used. They are spatial domain and transfer domain. But transfer domain techniques are more used compared to spatial domain.

### 3.1.1 Transfer Domain Techniques:

In this technique the coefficients of transfer domain are modified of Digital Image not like as the pixels values which is changed in spatial domain. Reverse process will be used to extract the watermark from watermarked image.

Some of the main transfer Domain techniques are:

   I.   Discrete Cosine Transform
   II.  Discrete Wavelet Transform
   III. Discrete Fourier Transform

Anyone can use individual transform techniques for watermarking but recently combination of these techniques are also used by researchers. By these combinations developers can used best features of any individual technique.

### 3.2 Applications of watermarking

Watermarking technologies is applied in every digital media whereas security and owner identification is needed[3]. A few most common applications are listed hereby.

#### 3.2.1    Owner Identification

The application of watermarking to which he developed is to identify the owner of any media. Some paper watermark is easily removed by some small exercise of attackers. So the digital watermark was introduced. In that the watermark is the internal part of digital media so that it cannot be easily detected and removed.

#### 3.2.2    Copy Protection

Illegal copying is also prevent by watermarking with copy protect bit. This protection requires copying devices to be integrated with the watermark detecting circuitry.

#### 3.2.3    Broadcast Monitoring

Broadcasting of TV channels and radio news is also monitoring by watermarking. It is generally done with the Paid media like sports broadcast or news broadcast.

#### 3.2.4    Medical applications

Medical media and documents also digitally verified, having the information of patient and the visiting doctors. These watermarks can be both visible and invisible. This watermarking helps doctors and medical applications to verify that the reports are not edited by illegal means.

#### 3.2.5    Fingerprinting

A fingerprinting is a technique by which a work can be assigned a unique identification by storing some digital information in it in the form of watermark. Detecting the watermark from any illegal copy can lead to the identification of the person who has leaked the original content. In cinema halls the movies are played digitally through satellite which has the watermark having theater identification so if theater identification detected from a pirated copy then action against a theater can be taken.

#### 3.2.6    Data Authentication

Authentication is the process of identify that the received content or data should be exact as it was sent. There should be no tampering done with it. So for that purpose sender embedded the digital watermark with the host data and it would be extracted at the receivers end and verified. Example like as CRC (cyclic redundancy check) or parity check.

## IV.    PERFORMANCE EVALUATION & THREATS

### 4.1 Threats for watermarking

There are so many threats for watermarking (Image) by which this process needs protected every time. As the watermarking techniques developed by researchers, hackers are developed new methods to attacks to destroy watermark. So every time algorithms need to be more robust for preventing attacks [4].

A few of the more obvious attacks are [5]:

- Image Compression - Lossy compression can result in the destruction of an image's watermark.
- Geometric transformations - the rotation, translation, sheering, or resizing of an image.
- Image Enhancements - Sharpening, colour calibration, contrast change.
- Image Composition - The addition of text, windowing with another image, etc.
- Information Reduction - Cropping
- Image filtering and the introduction of noise.
- Digital-to-analog conversion

In addition some sophisticated attacks are:

- Multiple watermarking – add second watermark to image that creates a problem of validating the owner information.
- Collusion attacks - Multiple receiving of the same host image.
- Forgery - Multiple recipients of different images, all with the same watermark (presumably here

identifying the owner), could be able to insert that same watermark into other images without the consent of the identified party.

These are the main threats or attacks for a watermarking are needed to be considered when an algorithm is chosen by user.

*4.2 Performance evaluation of watermarking algorithms*

Performance evaluation is very important part in the any algorithmic design in watermarking. The main task of this is to evaluate the quality matrices of algorithm or method to find out, how much he is effective?
Some of the quality matrices an image watermarking method or algorithm.

*4.2.1. Mean square error (MSE)*

The mean squared error (MSE) in an image watermarking is to estimate or measures the average of the squares of the "errors", between host image and watermark image [5].

$$MSE = 1 \div MN \sum_{i}^{M} \sum_{j}^{N} (Wij - Hij)^2$$

Where M, N is pixel values in host image
Wij= Pixel value in Watermarked Image
Hij= Pixel value in Host Image

*4.2.2 Peak signal to noise ratio (PSNR)*

PSNR (Peak Signal to Noise Ratio) is used to determine the Efficiency of Watermarking with respect to the noise. The noise will degrade the quality of image. The visual quality of watermarked and attacked images is measured using the Peak Signal to Noise Ratio [5]. It is given by

$$PSNR = 10 * \log (P^2/MSE)$$

Where p= maximum value in host image.

Imperceptibility of image is determined by this factor. More the PSNR shows that Watermarked image is perceptible or watermark is not recognized by naked eyes.